\documentclass[sn-basic]{sn-jnl}% Basic Springer Nature Reference Style/Chemistry Reference Style SI
\usepackage{graphicx}%
\usepackage{multirow}%
\usepackage{amsmath,amssymb,amsfonts}%
\usepackage{amsthm}%
\usepackage{mathrsfs}%
\usepackage[title]{appendix}%
\usepackage{xcolor}%
\usepackage{textcomp}%
\usepackage{manyfoot}%
\usepackage{booktabs}%
\usepackage{algorithm}%
\usepackage{algorithmicx}%
\usepackage{algpseudocode}%
\usepackage{listings}%
\usepackage[absolute,overlay]{textpos}%
\usepackage{bm}%
\usepackage[english]{babel}
\usepackage{array}
\usepackage{bbm}
\usepackage{float}
\theoremstyle{thmstyleone}%
\theoremstyle{thmstyletwo}%

\theoremstyle{thmstylethree}%

\begin{document}

\title[A bivariate spatial extreme mixture model for unreplicated heavy metal soil contamination]{A bivariate spatial extreme mixture model for unreplicated heavy metal soil contamination}


\author*[1]{\fnm{M. Daniela} \sur{Cuba}}\email{m.cuba.1@research.gla.ac.uk}

\author[1]{\fnm{Marian} \sur{Scott}}\email{Marian.Scott@glasgow.ac.uk}

\author[2]{\fnm{Benjamin P. } \sur{Marchant}}\email{benmarch@bgs.ac.uk}

\author[1]{\fnm{Daniela} \sur{Castro-Camilo}}\email{daniela.castrocamilo@glasgow.ac.uk}

\affil[1]{\orgdiv{School of Mathematics and Statistics}, \orgname{University of Glasgow}, \country{UK}}
\affil[2]{\orgname{British Geological Survey}, \country{Keyworth, Nottingham, UK}}


\abstract{Geostatistical models for multivariate applications such as heavy metal soil contamination work under Gaussian assumptions and may result in underestimated extreme values and misleading risk assessments \citep{marchant_spatial_2011}. A more suitable framework to analyse extreme values is extreme value theory (EVT). However, EVT relies on replications in time, which are generally not available in geochemical datasets. Therefore, using EVT {to map soil contamination} requires adaptation to {be used in} the {usual} single-replicate data {framework} of soil surveys. We propose {a bivariate spatial extreme} mixture model to model the body and tail of contaminant pairs, where the tails are {described} using a stationary generalised Pareto distribution. We demonstrate the performance of our model using a simulation study and through modelling bivariate soil contamination in the Glasgow conurbation. 
{Model results are given as maps of {predicted marginal concentrations} and probabilities of joint exceedance of soil guideline values. Marginal concentration maps show areas of elevated lead levels along the Clyde River and elevated levels of chromium around {the} south and southeast villages such as East Kilbride and Wishaw. The joint probability maps show higher probabilities of joint exceedance to the south and southeast of the city centre, following known legacy contamination regions in the Clyde River basin. }} 



% \pacs[MSC Classification]{35A01, 65L10, 65L12, 65L20, 65L70}
% \pacs[MSC Classification]{62G08, 62G32, 62H12, 62H25}

\maketitle

\section{Introduction}~\label{sec:intro}
Heavy metal soil contamination is the above-baseline accumulation of heavy metals (HMs) in the soil \citep{tang_diagnosis_2019, su_review_2014, mishra_heavy_2019}. The most common biological toxic HM elements are mercury (Hg), cadmium (Cd), lead (Pb), chromium (Cr), arsenic (As), zinc (Zn), copper (Cu), nickel (Ni), stannum (Sn, technical name for tin), and vanadium (V). While trace amounts of HMs are indispensable in the environment, acute and chronic exposure via direct contact, or through the digestive and respiratory tracts, pose a significant risk to public health \citep{oosthuizen_heavy_2012}. 

The main pathways for excessive HM accumulation in urban and rural environments are often anthropogenic \citep{gomez-sagasti_microbial_2012}. In urban areas,  primary sources include industrial waste and residue, chemical manufacturing, sewage, atmospheric deposition, and combustion of fossil fuels \citep{tang_diagnosis_2019, kupka_multiple_2021}. HM soil contamination is generally characterised by its wide spatial distribution, strong latency, irreversibility, and complex multivariate nature \citep{su_review_2014}. The remediation of contaminated soils is relatively slow compared to the remediation of contaminated water or air \citep{kupka_multiple_2021}. For this reason, understanding the extent and intensity of the contamination, particularly in urban and densely-populated areas, is essential to establish preventive public health measures and mitigate the impacts of soil HM contamination \citep{gomez-sagasti_microbial_2012}.  

HM contamination is usually mapped using data from soil surveys \citep{toth_maps_2016}, which use spatial sampling schemes, prioritising the spatial distribution as dictated by each contaminant's properties and other environmental factors \citep{khlifi_head_2010}. Concentration values are commonly modelled by performing a Box-Cox transformation (typically log transformation) {followed by a Gaussian process regression model such as trans-gaussian kriging} \citep{diggle_model-based_2007, lado_heavy_2008,lv_identifying_2015}. However, HM contamination distributions are known to be heavy-tailed, a property which persists even after transformation and is not captured using the Gaussian framework \citep{marchant_spatial_2011, marchant_robust_2010}. 
% \dcuba{In some cases, the persisting extreme values in the tail are considered outliers and removed before analysis. 
% However, not all extreme values are outliers and thus cannot be removed, resulting in distributions with longer tails and violating assumptions of the normality}. 
As a result, Gaussian models underestimate extreme values at the tails, which may result in an underestimation of the risk of public exposure to high levels of HM contamination.
% \dcuba{One way to check the lack of fit of Gaussian models, is through the evaluation of the distribution's kurtosis. 
% Values of the kurtosis higher than those of the Gaussian distribution would indicate the need for non-Gaussian distributions.}

% The estimated uncertainty for the kurtosis of the normal distribution is easily estimated as a function of sample size. The tail and extreme values of leptokurtic distributions, those with higher kurtosis than a Gaussian distribution, can therefore be considered not Gaussian and alternative modelling approaches should be pursued.}

% \dcuba{Marian wonders if we need the standard EVT theory (in reference to this paragraph).}
Extreme Value Theory (EVT) is the natural statistical framework for the analysis of extreme values. Recent years have seen a growing interest in EVT for environmental applications such as climate science or hydrology \citep{toulemonde_applications_2015}, motivating theoretical extensions to model extremes in spatial and multivariate settings.  Classical EVT distributions define extreme values as the maximum value inside a block (block-maxima) or values exceeding a given threshold (threshold exceedances). For the definition of the block-maxima approach, let $\{X_1,X_2,\cdots,X_n\}$ be a set of $n$ iid continuous random variables and $M_n = \text{max}\{X_1,\cdots,X_n\}$ represent the maxima. The extremal types theorem \citep{coles_introduction_2001} states that if there exist sequences $\{a_n>0\}$ and $\{b_n\}$ such that the normalisation of $M_n$ as $M^{*}_n = \frac{M_n-b_n}{a_n}$ converges to a non-degenerate function $G$ as $n \rightarrow \infty$, then $G$ belongs to the family of generalised extreme value distributions (GEV)  \citep{fisher_limiting_1928,gnedenko_sur_1943,mises_distribution_1954}. 
Threshold exceedances, also known as the peak-over-threshold approach (POT), represent an alternative to the block-maxima approach and are defined as observations of $X$ that exceed a given threshold $u$, i.e., $X-u|X>u$. If the conditions for the limiting characterisation of $M^{*}_n$ given above hold, threshold exceedances converge to the generalised Pareto distribution (GPD) when $u\to\infty$, with distribution function characterised by a scale $\sigma>0$ and a shape $\xi\in\mathbb{R}$ parameter, and given by
\begin{equation}\label{eq:GPD}
    H(y) = 1-\Big(1+\frac{\xi y}{\sigma}\Big)^{-1/\xi}.
\end{equation}
%They have a similar asymptotic characterisation, converging onto the generalised Pareto distribution (GPD) when $u\to\infty$ if the conditions for the limiting characterisation of $M^{*}_n$ given above hold \citep{james_pickands_iii_statistical_1975}.

% \dcc{Lines are needed here to highlight the main challenges of our application: 1) we can't use EVT as it is because of a lack of temporal replications; 2) we have a multivariate/bivariate spatial problem}
% , which is associated to the GEV through a point-process representation. 
% However, EVT models rely on replications in time at every sampling location, which are generally unavailable in geochemical surveys.  
An essential requirement for EVT is that of replications in time. In environmental applications, temporal replicates are generally available for phenomena with a temporal dimension. However, soil surveys are generally unreplicated, meaning replications in time are rarely available and preventing the use of EVT for applications such as mapping HM contamination. Additionally, the multivariate nature of HM contamination shows the presence of concomitant extremes, {which may be of special interest for public health planning due to the complexity posed by the risks associated with multiple contaminants}. It is clear then that a suitable spatial statistical modelling of HM contamination requires the use of multivariate spatial models that do not underestimate the marginal extreme behaviours, are able to capture extremal dependence between contaminants across space, and can handle unreplicated data. 

We propose to fill the gap between classical EVT based on replications and unreplicated multivariate spatial settings by using a continuous mixture of non-extreme and extreme distributions. Our methodology is proposed for the bivariate case where two contaminants are modelled simultaneously, accounting for the extremal dependence between variables through a coregionalisation framework. We assume the distribution of each contaminant can be decomposed into two components, representing the body and the tail of the distribution. The body of the distribution is composed of a combination of natural pedogenic processes and diffuse background contamination and represents the majority of the observations \citep{ander_methodology_2013}. The tail contains the extreme concentrations from contaminating anthropogenic or natural processes \citep{marchant_robust_2010}. Each component is modelled using suitable distributions for extreme and non-extreme concentrations and woven together
% \orange{Each component is generated by a process characterised by different distributions}\dcc{not sure what you mean here},
using a continuous mixture model representation. In principle, the body and tail processes of two contaminants can be differently affected by the same spatial factors, so we allow the mixture components to share spatial effects across variables inside a coregionalisation framework. Inference on the model is performed using the integrated nested Laplace approximation (INLA; \cite{rue_approximate_2009}), which allows Bayesian inference for the class of latent Gaussian models (LGMs) and can be fitted in \texttt{R} using the package \texttt{R-INLA}. Coregionalisation models are straightforward to fit with \texttt{R-INLA} \citep{krainski_advanced_2018}; however, mixture models lack the LGM representation needed for INLA, so we adapt a conditional approach following \citet{gomez-rubio_markov_2017}.
% , who showed that it is possible to condition the mixture model on fixed parameters and fit it using an INLA within a Monte Carlo Markov Chain (MCMC) framework. 
We assess our model using a simulation study and present a case study using data from the Geochemical Baseline Survey of the Environment (G-BASE) in the Glasgow Conurbation, where we show that our model correctly captures the tail behaviour of a bivariate geochemical dataset. An important byproduct of our modelling approach are risk maps showing {probabilities of two contaminants jointly exceeding their respective safety values as defined by the soil guideline values (SGV; \citealt{cole_using_2009})}. % We follow the workflow proposed by \citet{berild_importance_2022}, who extended the INLA within MCMC methodology by using importance sampling (IS) and Adaptive multiple importance sampling (AMIS) instead of the classical Metropolis-Hastings (MH) in MCMC as in \citet{gomez-rubio_markov_2017}, resulting in a significantly faster and computationally efficient model. 

The remainder of this manuscript proceeds as follows. Section \ref{sec:Background} provides an overview of the necessary statistical background. The coregionalised mixture model we propose is presented in Section \ref{sec:MixCoreg}. In Section \ref{sec:simstudy}, we detail the simulation study performed to assess the performance of our proposed model and present the results. Section \ref{sec:casestudy} {presents the results of the model fitted to chromium and lead as contaminant pairs in the Glasgow Conurbation, showing marginal concentrations and probability plots of joint exceedance of SGVs}. Finally, Section \ref{sec:C&D} provides a discussion. Supporting plots and tables are given in the Appendix.

%============================== Data description ==============================
\section{Statistical background}\label{sec:Background}

\subsection{Latent Gaussian models} \label{sec:LGM}
% The integrated nested Laplace approximation (INLA) is a probabilistic Bayesian inference framework for latent Gaussian models that provides approximations for the marginal posterior distributions for all parameters in the model \citep{rue_approximate_2009}. In practice, the implementation is relatively straightforward by using the R-INLA package in R. 
Latent Gaussian models (LGM) represent a wide subclass of structured additive regression models \citep{fahrmeir_multivariate_2001}. Let $\bm{y}=(y_1,...,y_n)$ represent the response vector with an assumed distribution in the exponential family of distributions with density function $\pi(y|\bm{\mathcal{X}},\bm{\theta})$ where $\bm{\theta}$ denotes a vector of hyperparameters. Using a link function $h(\cdot)$, $\bm{y}$ can be linked to covariates $\bm{Z} = \{\bm{X},\bm{U}\}$ through linear predictors
\begin{equation*}
    \bm{\eta} = \beta_0{\bm{1}} + \bm{\beta}\bm{X} + \sum_{k=1}^{K} f^{(k)}(\bm{u}_k),
\end{equation*}
where $\beta_0$ is an intercept, $\bm{\beta}$ is a vector of coefficients representing the linear effects of $\bm{X}$ on $\bm{\eta}$, and $f$ are unknown non-linear functions of covariates $\bm{U}$. 
A latent Gaussian model (LGM) arises in the case where a Gaussian prior is assumed for the latent field $\bm{\mathcal{X}} = \{\beta_0,\bm{\beta}, \bm{f}\}$. 
% Let $\mu_i$ denote a parameter of interest which is linked to a structured additive predictor, known as a linear predictor $\eta_i$, through some link function $g(\cdot)$ so that $g(\mu_i) = \eta_i$. Covariate information can then be included using various effects in an additive manner:
% \begin{equation*}
%     \eta_i = g(\mu_i) = \alpha + \sum_{k=1}^{n_\beta} \beta_k z_{ki} + \sum_{j=1}^{n_f} f^{(j)}(x_{ji}),
% \end{equation*}
% where $\alpha$ is an intercept, $\bm{z}_k$ are covariates with linear coefficients $\beta_k$, and $f^{(j)}(x_{ji})$ are unknown non-linear functions of the covariates $x_{j}$. 
% All components of the linear predictor are assumed to follow Gaussian priors. The latent Gaussian field $\bm{x}$ is defined as $\bm{x}=(\eta,\alpha,\boldsymbol{\beta},f^{(1)}, f^{(2)},...)$ \dcc{this has changed in the new INLA, check}.
A Bayesian hierarchical structure can be easily defined for the LGM, as
\begin{equation}
    \begin{aligned}
        \bm{y}|\bm{\mathcal{X}}, \bm{\theta}_1 & \sim \prod_{i=1}^n \pi(y_i|\bm{\mathcal{X}},\bm{\theta}_1)\\
        \bm{\mathcal{X}}|\bm{\theta}_2  & \sim \text{N}(\bm(0),\bm{Q}^{-1}_\pi(\bm{\theta}_2))\\
        \bm{\theta} = \{\bm{\theta}_1,\bm{\theta}_2\} & \sim \pi(\bm{\theta}),
    \end{aligned}
\end{equation}
where $\bm{\theta}$ are hyperparameters from the likelihood ($\bm{\theta}_1$) and the latent prior ($\bm{\theta}_2$). 

The aim of inference for LGM is the estimation of $\bm{\mathcal{X}} = \{\beta_0,\bm{\beta},\bm{f}\}$, the latent field, for which we need the marginal posteriors $\pi(\theta_j|\bm{y})$ and $\pi(\mathcal{X}_j|\bm{y})$. In its Bayesian hierarchical representation, it is possible to use Bayesian inference techniques such as MCMC to perform inference on LGMs. However, these techniques are often computationally unfeasible for large datasets \citep{rue_approximate_2009}. The integrated nested Laplace approximations (INLA) framework was proposed by \citet{rue_approximate_2009} as a computationally efficient inference framework for LGMs. In the original formulation of INLA, the linear predictors $\bm{\eta}$ were included in an augmented latent field 
\begin{equation*}
    \bm{\mathcal{X}} = \{\bm{\eta}, \beta_0, \bm{\beta}, \bm{f}\}. 
\end{equation*}
However, this results in a singular covariance matrix of $\bm{\mathcal{X}}$, requiring the addition of a Gaussian noise term in $\bm{\eta}$ as
\begin{equation}
    \bm{\eta} = \beta_0\bm{1} + \bm{\beta}\bm{X} + \sum_{k=1}^{k}f^{k}(\bm{u}_k) + \bm{\epsilon},
\end{equation}
where $\bm{\epsilon} \sim (\bm{0},\tau^{-1}\bm{1}$, with a large but fixed $\tau$ \citep{van_niekerk_new_2023}. This enhancement is no longer necessary in the new formulation of INLA. Here, the latent field is defined as before
\begin{equation*}
    \bm{\mathcal{X}} = \{\beta_0, \bm{\beta},\bm{f}\},
\end{equation*}
with the Gaussian prior $\bm{\mathcal{X}}|\bm{\theta} \sim \text{N}(\bm{0},\bm{Q}^{-1}(\bm{\theta}))$. 
The linear predictors are defined as 
\begin{equation*}
    \bm{\eta} = \bm{A}\bm{\mathcal{X}},
\end{equation*}
where $\bm{A}$ is a sparse design matrix linking the linear predictors to the latent field \citep{van_niekerk_new_2023}. 

The Gaussian approximation $\pi_G(\bm{\mathcal{X}}|\bm{\theta}, \bm{y})$ to $\pi(\bm{\mathcal{X}}|\bm{\theta}, \bm{y})$ is calculated from a second order expansion of the likelihood around the mode
\citep{van_niekerk_new_2023} resulting in 
\begin{equation*}
    \bm{\mathcal{X}}|\bm{\theta},\bm{y} \sim \text{N}\left(\bm{\mu}(\bm{\theta}),\bm{Q}{-1}_{\bm{\mathcal{X}}}(\bm{\theta})\right),
\end{equation*}
where $\bm{Q}_{\bm{\mathcal{X}}}^{-1}$ represents a precision matrix. The univariate conditional posteriors follow naturally from the joint Gaussian approximation as 
\begin{equation}\label{eq:condX}
    \mathcal{X}_j|\bm{\theta},\bm{y} \sim \text{N}\left(\bm{\mu}(\bm{\theta})_j, \left(\bm{Q}^{-1}_{\mathcal{X}}\right)_{jj}\right).
\end{equation}
The marginal posterior $\Tilde{\pi}(\mathcal{X}_j|\bm{y})$ can be obtained by integrating $\theta$ out of \ref{eq:condX} by using $K$ points of integration, $\theta_k$ \citep{van_niekerk_new_2023}.

The marginal posteriors of the linear predictors $\Tilde{\pi}(\eta_i|\bm{y})$ were automatically calculated in the original INLA formulation of \citet{rue_approximate_2009}. However, the new formulation requires first calculating $\Tilde{\pi}(\eta_i|\bm{\theta},\bm{y})$. The resulting posterior means of $\bm{\eta}$ and  $\bm{\mathcal{X}}$, based on the Gaussian assumption of the conditional posterior, can be inaccurate \citep{van_niekerk_new_2023}. A Variational Bayes correction to the means, proposed by \citet{van_niekerk_low-rank_2021}, can improve the posterior means of the latent field. The corrected posterior inference fo $\bm{\eta}$ is
\begin{equation*}
    \begin{aligned}
        \eta_j \mid \boldsymbol{\theta}, \boldsymbol{y} & \sim N\left(\mu_j(\boldsymbol{\theta}), \sigma_j^2(\boldsymbol{\theta})\right) \\
        \mu_j(\boldsymbol{\theta}) & =(\boldsymbol{A} \boldsymbol{\mu}^*(\boldsymbol{\theta}))_j \\
        % \sigma_j^2(\boldsymbol{\theta}) & =\sum_{i l} A_{j i} A_{j l} C_{i l} \\
        \tilde{\pi}\left(\eta_j \mid \boldsymbol{y}\right) & \approx \sum_{k=1}^K \pi_G\left(\eta_j \mid \boldsymbol{\theta}_k, \boldsymbol{y}\right) \tilde{\pi}\left(\boldsymbol{\theta}_k \mid \boldsymbol{y}\right) \delta_k,
    \end{aligned}
\end{equation*}
where $\delta_k$ are the weights of the integration points $\theta_k$ when computing the marginal posterior of $\mathcal{X}_j$. 

\subsection{Mixture models}\label{sec:Mix}
A mixture model is a convex, weighted combination of multiple distributions representing the different assumed underlying groups in the data. An observation $y_i$ is assumed to come from one of $k$ data generating processes, each with their own set of parameters \citep{mclachlan_finite_2019}, i.e.,  
\begin{equation}\label{eq:mix}
    y_i \sim \sum_{k=1}^K w_k f_k(y_i,\theta_k),
\end{equation}
where ${f_1(\cdot|\theta_1),\ldots,f_K(\cdot|\theta_K)}$ is a set of parametric distributions (one for each latent group in the data), and $\bm{w}=(w_1,...,w_K)$ are their associated weights with $\sum_{k=1}^K w_k = 1$. 

Unless we assume the mixture weights are known, mixture models do not have a latent Gaussian representation. Consequently, inference cannot be carried out in INLA without adjustment \citep{gomez-rubio_mixture_2017}, as in the \textit{augmented data} approach in \citet{dempster_maximum_1977}. This approach considers an auxiliary variable $\textbf{z}=(z_1,...,z_n)$ that assigns each observation to a group in the set ${1,...,K}$. The mixture model can then be represented as 
\begin{equation*}
    y_i | z_i \sim f_{z_i}(y_i|\theta_{z_i}), \quad z_i \in {1,...,K}. 
\end{equation*}
The distribution of $\bm{z}$ can be stated as 
\begin{equation*}
    \pi(z_i = k|\bm{w}) = w_k, \quad k=1,...,K,
\end{equation*}
resulting in the likelihood
\begin{equation*}
    \pi(\bm{y},\bm{z}|\boldsymbol{\theta},\bm{w}) = \Big(\prod_{i=1}^n f_{z_i}(y_i|\theta_{z_i})\Big)w_{k}^{n_k},
\end{equation*}
where $n_k$ is the number of observations in group $k$. When conditioning on $\bm{z}$, under the assumption that the groups are independent, the mixture model has a latent Gaussian representation and can be fitted with INLA using various possible approaches such as \citet{gomez-rubio_mixture_2017} and \citet{gomez-rubio_markov_2017}. \citet{gomez-rubio_mixture_2017} shows that the posterior marginal of the conditional mixture model can be represented as 
\begin{equation*}
    \pi(\theta_t|\bm{y}) = \sum_{z\in \mathcal{Z}}\pi(\theta_t|\bm{y},\bm{z} = z)\pi(\bm{z}=z|\bm{y}),\quad t=1,...,\text{dim}(\boldsymbol{\theta}). 
\end{equation*}
The above requires the posterior distribution of $\bm{z}$, which can be estimated using the INLA within Monte Carlo Markov Chain (MCMC) approach in \citet{gomez-rubio_markov_2017}. 
The adjustment to produce the latent Gaussian representation requires conditioning on fixed parameters or hyperparameters of $\bm{z}$, denoted as $\bm{z}_c$. Denoting the non-fixed parameters by $\bm{z}_{(-c)}$, the resulting posterior can be expressed as
\begin{equation*}
    \pi(\bm{z}|\bm{y}) \propto \pi(\bm{y}|\bm{z}_{-c})\pi(\bm{z}_{-c}|\bm{z}_{c})\pi(\bm{z}_{c}),
\end{equation*}
where $\pi(\bm{y}|\bm{z}_{-c})\pi(\bm{z}_{-c}|\bm{z}_{c})$ is a LGM suitable for INLA. Although it is possible to perform inference on $\bm{z}_c$ using MCMC or importance sampling \citep{berild_importance_2022}, we choose a fixed \textit{a priori} value for $\bm{z}_c$ due to computational limitations. The choice of $\bm{z}_c$ is done using an extensive grid search of possible values and model-performance metrics to assess optimum values. 

\subsection{Coregionalisation framework} \label{sec:Coreg}
Multivariate response models with latent Gaussian characterisations can be fitted inside the coregionalisation framework \citep{krainski_advanced_2018}. The framework allows for a different likelihood function for each variable and models the dependence structure between variables by introducing shared components in the linear predictor. In the bivariate case, let $\bm{y}_1(\bm{s})$ and $\bm{y}_2 (\bm{s})$ be two spatial variables for locations $\bm{s} \in \mathcal{S}$. Simple linear predictors for a coregionalisation model can be defined as 
\begin{equation}\label{eq:coreg}
\begin{aligned}
   \eta_1(\bm{s}) & = \alpha_1 + z_1(\bm{s})\\
   \eta_2(\bm{s}) & = \alpha_2 + \lambda z_1(\bm{s}) + z_2(\bm{s}),
    \end{aligned}
\end{equation}
 where $\eta_1(\bm{s})$ and $\eta_2(\bm{s})$ are the linear predictors at $\bm{s}$ of $\bm{y}_1(\bm{s})$ and $\bm{y}_2 (\bm{s})$, respectively, $\alpha_1$ and $ \alpha_2$ are intercepts, $z_1(\bm{s})$ and $z_2(\bm{s})$ are spatial random effects \citep{lindgren_explicit_2011}, and $\lambda$ is a weight for the shared spatial effect \citep{krainski_advanced_2018}. In essence, the linear predictors of $\bm{y}_1$ and $\bm{y}_2$ are dependent on a shared component, $z_1(\bm{s})$, which captures dependence between variables. While $\bm{y}_1$ has a single spatial effects term, $\bm{y}_2$ can have a second spatial effects term ($z_2$), adding flexibility to the model. Inference on the model is easily performed in INLA \citep{krainski_advanced_2018}, using two likelihoods. 
 It can be extended beyond the bivariate case to include different likelihoods for each response variable, but the implementation is restricted due to the tradeoff between accuracy and computational costs. Additionally, the linear predictors can also contain linear and non-linear effects, as well as more shared components of different forms.
 
\section{Proposed methodology and inference} \label{sec:MixCoreg}
Our bivariate spatial extreme mixture model for unreplicated data combines bivariate mixture models for an accurate representation of the body and tail of the distribution of each contaminant while a coregionalisation structure incorporates the spatial dependencies within a latent Gaussian model framework.
Specifically, we construct two spatial mixture models with $K=2$ mixing components for variables $y_1(\bm{s})$ and $y_2(\bm{s})$ at locations $\bm{s} \in \mathcal{S}$. 
%where $\bm{s}_{T_1}$ and $\bm{s}_{T_2}$ are the respective locations where variables $y_1$ and $y_2$ are extreme and $\bm{s}_{B_1}$ and $\bm{s}_{B_2}$ are non-extreme locations.
The components account for the body and tail observations of each variable. The spatial mixture models are defined as
\begin{align}\label{eq:bivmix}
     y_1(\bm{s}) & \sim p_1f_{B_1}(y_1(\bm{s})|\theta_{B_1}) + (1-p_1)f_{T_1}(y_1(\bm{s})|\theta_{T_1}),\nonumber\\
     y_2(\bm{s}) & \sim p_2f_{B_2}(y_2(\bm{s})|\theta_{B_2}) + (1-p_2)f_{T_2}(y_2(\bm{s})|\theta_{T_2}),
\end{align}
where, for $i \in \{1,2\}$, $f_{B_i}$ is the density of the non-extreme observations in the body, $f_{T_i}$ is the density of the extremes in the tail, and $p_i$ is the mixing proportion or the probability that an observation belongs to the body of the distribution. 
From Section \ref{sec:Coreg}, the coregionalisation framework is defined through linear predictors with shared components.
For our mixture models in~\eqref{eq:bivmix}, we carefully explored the inclusion of shared components on the bulks and tails to account for dependence at non-extreme and extreme values, respectively. 
Even though the shared components are flexible and can be tailored for each application, components should be linked only when necessary since increasing the number of shared components considerably increases computational costs. 
Our coregionalised mixture model shares spatial components only in the tails to account for extremal dependence, while the non-extreme components have shared dependence only through common covariates. The model can be expressed as
\begin{equation}\label{eq:bivmodel}
    \begin{aligned}
    \eta_{B_1}(\bm{s}) & =\alpha_{B_1} + z_{B_1}(\bm{s})+ \sum_{j \in \mathcal{J}} \beta_{{B_{1j}}} x_j(\bm{s}), & \qquad\eta_{T_1}(\bm{s})  & =\alpha_{T_1} + z_{T_1}(\bm{s})+ \sum_{j \in \mathcal{J}} \beta_{{T_{1j}}} x_j(\bm{s}),\\
    \eta_{B_2}(\bm{s}) & =\alpha_{B_2} + z_{B_2}(\bm{s})+ \sum_{j \in \mathcal{J}} \beta_{{B_{2j}}} x_j(\bm{s}), & \qquad
    \eta_{T_2}(\bm{s})  &=\alpha_{T_2} + \lambda_T z_{T_1}(\bm{s}) + z_{T_2}(\bm{s})+ \sum_{j \in \mathcal{J}} \beta_{{T_{2j}}} x_j(\bm{s}),
\end{aligned}
\end{equation}
where for $i=1,2,$
$\eta_{B_i}(\textbf{s})$ and $\eta_{T_i}(\textbf{s})$ are the corresponding linear predictors for $f_{B_i}$ and $f_{T_i}$ respectively, 
$z_{B_i}$ and $z_{T_i}$ are random spatial effects, $x_j$ are covariates, $\alpha_{B_i}$ and $\alpha_{T_i}$ are intercepts, $\beta_{B_{ij}}$ and $\beta_{T_{ij}}$ are coefficients corresponding to the covariates $x_j$ for the body and tail respectively, and $\lambda_T$ is a scaling coefficient for the shared random spatial effect $z_{T_1}$.
Although different likelihoods can be assigned to each component (namely, $B_1$, $T_1$, $B_2$ and $T_2$), our specific application driven by the need to correctly characterise extreme and non-extreme HM concentrations, motivates the use of a Gaussian likelihood for the bodies $B_1$ and $B_2$, and a generalised Pareto distribution (GPD) for the tails $T_1$ and $T_2$. 
By using the GPD we are implicitly defining extreme observations as exceedances over suitably high thresholds $u_1$ and $u_2$ for $\bm{y}_1$ and $\bm{y}_2$, respectively. 

Inference for our model is based on the conditional latent Gaussian field framework proposed by \citet{gomez-rubio_markov_2017}, where we replace the MCMC inference with a simple conditional approach based on conditioning parameters \textit{a priori}, similar to the importance sampling approach to conditional mixture models by \citet{berild_importance_2022}. 
As noted in Section~\ref{sec:intro}, the GPD requires replicates over time.
% and perform inference using the stochastic partial differential equations (SPDE) approach in INLA \citep{lindgren_explicit_2011}. 
The implementation of the GPD likelihood in  INLA links the linear predictor to a fixed $\alpha$-quantile of the distribution, specifically, $q_{\alpha}(\textbf{s}) = \exp\{\eta_{T_i}(\textbf{s})\}$ with $P(y<q_{\alpha}(\bm{s}))=\alpha$. This parametrisation implicitly assumes replicates at each location $\textbf{s}\in\mathcal{S}$.
Interestingly, this is not the case for the Gaussian likelihood, where latent Gaussian models can be fitted following the usual geostatistical design of single replicates over space. 
To circumvent the need for replications of tail values, we assume spatial stationarity of threshold exceedances and model the tail of each HM contaminant using a stationary GPD as in~\eqref{eq:GPD}.
Then, we transform the tail values to a common Gaussian scale using the probability integral transform.
Finally, tail values in our {bivariate spatial extreme mixture model} are modelled using the alternative tail distribution density
%The parameter is related to the linear predictor through $\sigma = \frac{\xi\text{exp}(\eta)}{(1-\alpha)^{-\xi}-1}$ under the expectation of replicates at each location.
% It results in a unique distribution (parameter set) at each location as $\sigma(\textbf{s})$ but common shape parameter $\xi$, depending on the $\alpha$-quantile being modelled. 
%Under the assumption of spatial stationarity of the exceedances of the thresholds $u_1$ and $u_2$ (stationary $f_{T_1}$ and $f_{T_2}$), the tail can be modelled using a GPD with distribution function
%\begin{equation}\label{eq:GPD}
%    H(y) = 1-\Big(1+\frac{\xi (y-\mu)}{\sigma}\Big)^{-1/\xi},
%\end{equation}
%where $\mu$ is the location parameter, $\sigma>0$ is the scale parameter, and $\xi>0$ is the shape parameter. 
%A transformation is required to a Gaussian distribution to circumvent the problem of lack of replications using
\begin{equation}\label{eq:modtail}
    f'_{T_i}(\bm{s}) = \Phi(F_{T_i}^{-1}(y_{i}(\bm{s});0,\sigma, \xi)), \quad y_i(\bm{s})>0
\end{equation} 
where $y_{T_i}(\bm{s})$ represents the extreme values constituting the tail of contaminant, $\Phi$ is the standard normal distribution and $F_{T_i}$ is the GPD fitted to all exceedances of the threshold $u_i$ for variable $y_i$. {As per the transformation in ~\eqref{eq:modtail}, the two distributional components of the coregionalisation model are Gaussian and {non-informative} Gaussian priors are assigned to all components of the linear predictors}.
Pointwise predictions and their uncertainties can be obtained using posterior samples, with the necessary back-transformation for values belonging to the tail.

\begin{figure}
    \centering
    \includegraphics[scale=0.45]{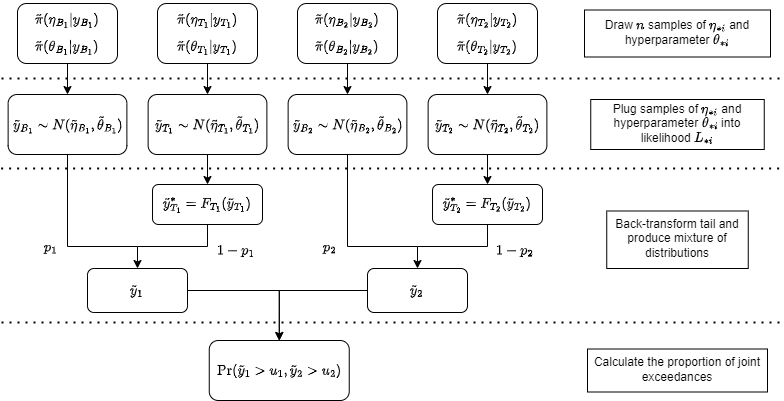}
    \caption{{Flow chart of the Monte Carlo method used to sample from posterior predictive distributions of $y_1$ and $y_2$ and obtain a map of the probability of joint exceedance of threshold $u_1$ and $u_2$ respectively. The process is repeated 1000 times.}}
    \label{fig:mc_probs}
\end{figure}

For risk assessment purposes, we also derive joint exceedance probability maps.
Specifically, we compute $\text{Pr}(y_1(s_i)>u_1|y_2(s_i)>u_2)$, where $u_1$ and $u_2$ are sufficiently high threshold (following soil guideline values).
These values are obtained by sampling from the posterior predictive distribution, which can be done using a multi-step Monte Carlo method summarised in Figure \ref{fig:mc_probs}. It is first necessary to sample from the posteriors of the linear predictor and the hyperparameters. These samples are then used to obtain samples of the posterior predictive distribution of each component after a back-transformation of the tail distribution and mixing of the distributions according to the mixture proportions $p_1$ and $p_2$ as in \eqref{eq:bivmix}. Finally, the joint exceedance probabilities are computed empirically. 
% The number of posterior predictive samples of $\Tilde{y}_1$ and $\Tilde{y}_2$ that simultaneously exceed their respective thresholds, $u_1$ and $u_2$, is divided by the total number of samples to obtain the probabilities of joint exceedance given that the dependence between $\Tilde{y}_1$ and $\Tilde{y}_2$ is accounted for at the latent level. 
This process is repeated $1000$ to obtain measures of uncertainty of the Monte Carlo procedure.

 \newpage

\section{Simulation study}\label{sec:simstudy}
In this section, we show the construction and results of an extensive simulation study aimed at assessing the performance of the proposed {bivariate spatial extreme mixture model} for a single realisation mimicking HM soil contamination. {A total of four scenarios were considered, set up to mimic behaviour observed in real geochemical datasets. }

\subsection{Data-generating process}\label{sec:Data}
The data were simulated directly from the model in \eqref{eq:bivmodel} after the adjustment proposed in \eqref{eq:modtail}. We produced a total of $N=1000$ simulations of $n=1000$ observations over the region $\mathcal{S}=[0,100]^2$ for two response variables, $y_1$ and $y_2$. The random spatial effects are simulated as Gaussian Processes (GP) with Matern covariance function \citep{matern_spatial_1960} defined as 
\begin{equation*}
    C_\nu(d) = \sigma^2 \frac{2^{1-\nu}}{\Gamma(\nu)}\left(\sqrt{2\nu}\frac{d}{\rho}\right)^{\nu}\text{K}_\nu \left(\sqrt{2\nu}\frac{d}{\rho}\right),
\end{equation*}
where $d$ is the Euclidean distance between two observations, $\Gamma$ is the gamma function, $\text{K}_nu$ is the modified Bessel function of the second kind, $\rho$ is the range of spatial dependence, and $\nu=1$ is the smoothness parameter. Parameter values for the proposed simulation scenarios are given in Table \ref{tab:biv_parvalsAB}.

\begin{table}
\centering
\begin{tabular}{lll}
                               \hline
\textbf{Parameter}             & \textbf{A} & \textbf{B} \\ \hline
$\text{Variation 1}: (p_1, p_2)$       & (0.75, 0.75)              & (0.75, 0.75)              \\
$\text{Variation 2}: (p_1, p_2)$       & (0.9, 0.9)              &     (0.9, 0.9)          \\
$(\alpha_{B1}, \alpha_{T1})$       & (1,0)              & (1,0)              \\
$(\alpha_{B2}, \alpha_{T2})$       & (1,0)              & (1,0)              \\
% $(\tau_{T1}, \tau_{T2})$           & (0.01, 0.01)              & (0.01, 0.01)              \\
% $(\tau_{B2}, \tau_{T2})$           & (0.01, 0.01)              & (0.01, 0.01)              \\
% $(x_1,x_2)$                          & $\{U(0,1),U(0,1)\}$       & $\{U(0,1),U(0,1)\}$\\
$(\beta_{B1},\beta_{T1})$          & (0.1,0.25) & (0.1,0.25)\\
$(\beta_{B2},\beta_{T2})$          & (0.1,0.25)               & (0.1,0.25)               \\
$\lambda$     & 0.25        & 0.9\\
% 2: $\lambda$     & (0.25)         & (0.9)         \\
$(\rho_1, \rho_2)$             & (10, 15)            & (10, 15)\\ 
$(\sigma_{T1},\sigma_{T2})$ & (1,1) & (1,1)\\
$(\xi_1, \xi_2)$             & (0.05, 0.25)            & (0.5, 0.25) \\\hline  \\
\end{tabular}
\caption{Parameter values for the bivariate scenarios A and B, representing small and large spatial extremal dependence through the weight of the shared spatial effect $\lambda$. Each scenario is further subdivided into variations 1 and 2, resulting in four scenarios: A1, A2, B1, and B2. Variations 1 and 2 represent two mixture proportions, $p=0.75$ and $p=0.9$, respectively.}
% \dcc{see my comment above}. 

\label{tab:biv_parvalsAB}
\end{table}

\subsection{Classification} \label{sec:class}
The model requires \textit{a priori} classification of observations as belonging to the body or tail of the distribution. While the choice of mixture proportion $p$ also defines threshold $u$, given that the proportion of observations exceeding threshold $u$ is the same as $p$, setting all exceedances from $u$ as belonging to the tail results in an upper truncation for the body. As a result, we propose a classification based on the Metropolis-Hastings algorithm, which results in a soft boundary between body and tail. The classification is as follows
\begin{enumerate}
    \item For the body and tail distributions, $f_B(y(\bm{s})) \equiv \text{N}(\mu,\tau^2)$ and $f_T(y(\bm{s})) \equiv \text{GPD}(u_i, \sigma,\xi)$, respectively, estimate all parameters using maximum likelihood.
    \item For each observation in $y(\bm{s})$ compute the density under $f_B$ and under $f_Y$ as $p_B(y(s_j))$ and $p_T(y(s_j))$ respectively.
    \item Obtain the classification ratio $p_\alpha = \min\Big\{1,\frac{p_T(y(s_j))}{p_B(y(s_j))}\Big\}$.
    \item Draw a random sample from a uniform distribution $u_\alpha \sim \text{U}(0,1)$.
    \item Assign the observation $y(s_j)$ as belonging to the tail if $p_\alpha >= u_\alpha$.
    \item Repeat the process $n_c=100$ times for each observation and {assign the membership that appears the most number of times in the $n_c$ samples.}
\end{enumerate}
    
% \end{equation}
\subsection{Evaluation}\label{sec:sim}
Assessment is performed by quantifying the true parameter coverage probability (the probability that the true parameter value is in the confidence interval), as well as the parameter estimation error using root mean squared error (RMSE) computed as 
\begin{equation*}
    \text{RMSE} = \sqrt{\frac{\sum_{i=1}^{N}(\hat{y}_i - y_i)^2}{N}}.
\end{equation*}

Generally, a small RMSE indicates a good model fit, given that the predicted parameter values are close to the true parameter values. The assessment of the categorisation of observations as body or tail, a form of pre-processing, is performed using standard categorisation evaluation techniques such as computing the accuracy, sensitivity, specificity, and precision.

\subsection{Results}\label{sec:results}
\begin{figure}
    \centering
    \includegraphics[scale=0.35]{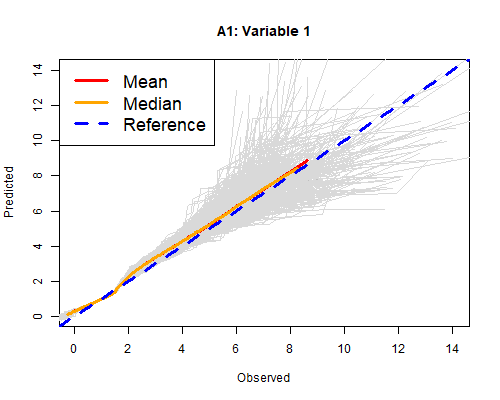}
    \includegraphics[scale=0.35]{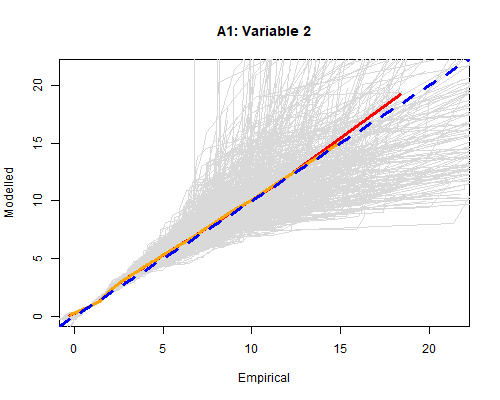}
    \caption{Q-Q plots of all simulations (grey) of bivariate scenario A1, for variables 1 and 2. The mean and median of the simulations are shown in red and orange, respectively, while the reference line is given in blue. }
    \label{fig:resultsA1}
\end{figure}

% Each simulation scenario described in Section \ref{sec:Data} was simulated $N = 500$ times. %, drawing 150 samples of the IS process for each simulation. 
% \subsubsection{Scenario A}
Figure \ref{fig:resultsA1} shows the Q-Q plots of the results for simulation A1. The figure shows that the model performs better for $y_1$ (variable 1), displaying smaller variability at higher values than $y_2$ (variable 2). However, both the median and the mean of the simulations for both variables accurately capture the true behaviour of the data for the body and tail of the distributions. Table \ref{tab:A1} shows the coverage probability for the parameters of the linear predictor, $\boldsymbol{\alpha}$ and $\boldsymbol{\beta}$.
We can see that they are well captured and have coverage probabilities greater than $0.9$, with the exception of $\alpha_{T_1}$, which might account for abrupt behaviour around the transition between body and tail. 
% \dcc{any explanation for this behaviour?}. 
Of the remaining parameters, only $\lambda$ has lower coverage probabilities, indicating the fit is not as good in $y_2$ as in $y_1$, as is expected given the asymmetric construction of the model. 

\begin{table}
\centering
\begin{tabular}{rlrrrrrr}
  \hline
 &\textbf{Parameter} &\textbf{True Val}& \textbf{Mean} & \textbf{Median} & \textbf{Sd} & \textbf{Coverage Pr} & \textbf{RMSE} \\
  \hline
 & $\alpha_{B1}$ & 1.00 & 1.04 & 1.04 & 0.01 & 0.99 & 0.04 \\ 
   & $\alpha_{T1}$ & 0.00 & -0.22 & -0.23 & 0.05 & 0.38 & 0.23 \\ 
 & $\alpha_{B2}$ & 1.00 & 1.05 & 1.04 & 0.03 & 0.99 & 0.06 \\ 
   & $\alpha_{T2}$ & 0.00 & -0.24 & -0.23 & 0.07 & 0.88 & 0.25 \\ 
   & $\beta_{B1_{1}}$ & 0.10 & 0.07 & 0.07 & 0.01 & 0.99 & 0.03 \\ 
   & $\beta_{B1_{2}}$ & 0.25 & 0.18 & 0.18 & 0.01 & 0.99 & 0.07 \\ 
   & $\beta_{T1_{1}}$ & 0.10 & 0.07 & 0.07 & 0.06 & 0.97 & 0.07 \\ 
   & $\beta_{T1_{2}}$ & 0.25 & 0.18 & 0.18 & 0.04 & 0.98 & 0.09 \\ 
   & $\beta_{B2_{1}}$ & 0.10 & 0.08 & 0.07 & 0.01 & 0.99 & 0.03 \\ 
   & $\beta_{B2_{2}}$ & 0.25 & 0.19 & 0.19 & 0.02 & 0.99 & 0.06 \\ 
   & $\beta_{T2_{1}}$ & 0.10 & 0.07 & 0.07 & 0.06 & 0.98 & 0.07 \\ 
   & $\beta_{T2_{2}}$ & 0.25 & 0.18 & 0.18 & 0.05 & 0.98 & 0.08 \\ 
   & $\tau_{1}$ & 0.01 & 0.20 & 0.19 & 0.10 & 0.74 & 0.22 \\ 
   & $\tau_{2}$ & 0.01 & 0.14 & 0.13 & 0.06 & 0.96 & 0.15 \\ 
   & $\rho_{T1}$ & 10.00 & 38.79 & 28.83 & 35.70 & 0.89 & 45.78 \\ 
   & $\rho_{T2}$ & 15.00 & 129.62 & 90.98 & 172.25 & 0.80 & 172.48 \\ 
   & $\lambda$\ & 0.25 & 0.97 & 0.97 & 0.08 & 0.32 & 0.73 \\ 
   & $\sigma_1$ & 1 & 1.004 & 0.99 & 0.14 & 0.92 & 0.14\\
   & $\sigma_2$ & 1 & 1.16 & 1.14 & 0.17 & 0.89 & 0.23 \\
& $\xi_1$\ & 0.05 & 0.06 & 0.06 & 0.11 & 0.93 & 0.11 \\ 
   & $\xi_2$\ & 0.25 & 0.27 & 0.27 & 0.12 & 0.95 & 0.94 \\ 
   \hline\\
\end{tabular}

\caption{Summary of results of A1. The table shows the parameter's true value; estimated parameter mean, median and standard deviation, 95\% coverage probability, and the mean RMSE. }
\label{tab:A1}
\end{table}

The results for A2 show a pattern similar to that of A1. The Q-Q plots in Figure \ref{fig:resultsA2} in the Appendix show a larger variability displayed in the tail of $y_2$ than in $y_1$. Additionally, $y_2$ is slightly underestimated at extreme values. Table \ref{tab:A2} in Appendix summarises the parameter estimates. Once again, the linear coefficients of $\eta$ are well captured by the model while $\lambda$ is consistently overestimated, similar to A1.

The performance assessments of the classification of observations as body or tail for A1 and A2 are given in Table \ref{tab:allA} in the Appendix. Carried out $\textit{a priori}$ with the method in Section \ref{sec:class}, the classification of both is approximately $\sim 90\%$ of the observations begin classified correctly.

Scenario B had a larger value for the weight of the shared spatial component, $\lambda = 0.9$, indicating stronger extremal dependence between variables. Figure \ref{fig:resultsB1} in the Appendix shows that for variable 1, the model performs as expected, similarly to results in A1 and A2. Variable 2 follows the pattern seen in A1 and A2, where variability is increased in the tail. However, in this scenario, the mean and median show a slight overestimation at the extremes. Table \ref{tab:B1} in Appendix shows the model has good coverage probabilities for most parameters. The large standard deviation of the range parameters, $\rho_1$ and $\rho_2$, show the model struggles to calculate the range of the spatial components, a possible explanation for the increased variability in the tails of the distribution. 

The results for B2 are shown in Figure \ref{fig:resultsB2} in the Appendix. The model performs similarly to B1, with the mean showing sensitivity to large values and a correct median for both variables. The summary of the estimated model parameters in Table \ref{tab:B2} show that the model correctly estimates most parameters, with the exception of $\rho_1$ and $\rho_2$, which experience an even bigger variability in estimation than previous scenarios. The coverage probability of the ranges, $\rho_1$ and $\rho_2$ seems to be lower for Variation 2, meaning the model struggles to accurately estimate the range when there are fewer extreme observations. 

The classification of the observations for B1 and B2 (Table \ref{tab:allB}) is similar to scenario A. The relatively lower specificity values indicate a poorer categorisation of the tail observations than the body. Overall, the results indicate the model did not suffer a loss of power with a larger imbalance in classes. 

\section{Case Study: Heavy Metal Soil Contamination Glasgow Conurbation}\label{sec:casestudy}
% \dcc{missing: info about priors} - described in methodology

The bivariate spatial mixture model was applied to the log-transformed concentrations of Cr and Pb in the Glasgow Conurbation using data from the Geochemical Baseline Survey of the Environment (G-BASE; see \citealt{johnson_g-base_2005} for details). The results of the model are compared to a standard Gaussian model with no mixture where both contaminants are fitted independently. 
% \dcuba{Furthermore, we use our mixture model to produce maps of joint exceedance probabilities, allowing for appropriate risk assessments of the Cr-Pb contamination in the area. }

\subsection{Data Description}
The British Geological Survey (BGS) performed the G-BASE survey from the 1960s to 2014 \citep{johnson_g-base_2005}. Initially commissioned for mineral exploration, it is a valuable tool for the systematic assessment of the geochemical background of the UK environment. The area of interest is the Glasgow Conurbation in the Clyde River Basin, west of Scotland, where the data consist of approximately $2745$ topsoil (5-20cm deep) samples taken at approximately $4$ observations per $\text{km} ^2$ in the urban areas and $1$ per $1 \text{ km} ^2$ in rural areas. Each sample was decomposed chemically using X-ray fluorescence and the concentration of each element in the sample in parts per million (ppm). Further details can be found in \citet{johnson_g-base_2005}. 

\begin{figure}
    \centering
    \includegraphics[scale=0.28]{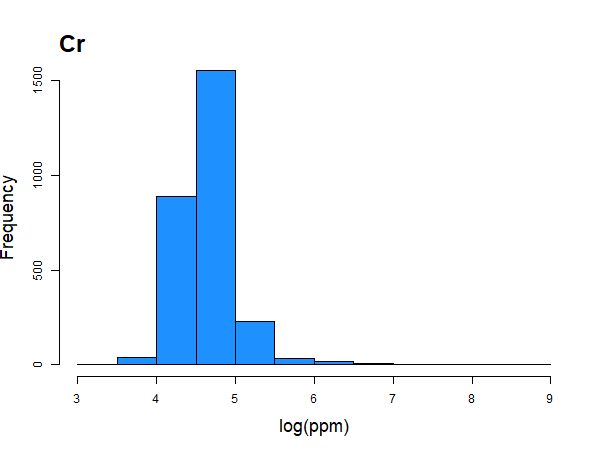}
    \includegraphics[scale=0.28]{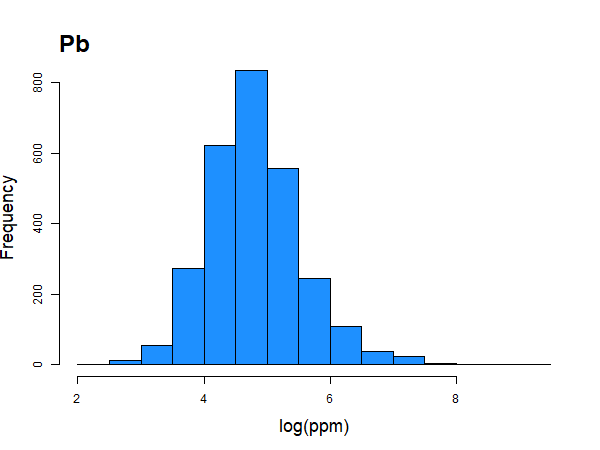}
    \caption{Histogram of Cr (left) and Pb (right) after log transformations.}
    \label{fig:hists}
\end{figure}

{As a pre-processing step, the data are first transformed using a log transformation. Figure \ref{fig:hists} shows histograms of the data post-transformation. Although the data appear approximately normal, kurtosis is 18.32 and 4.69 for Cr and Pb, respectively. The kurtosis shows that the tail of Cr is significantly heavier than that of a normal distribution, whereas Pb has a lighter tail that is just outside the expected values for a Gaussian distribution of the same sample size. This again highlights the need for an alternative approach to the usual Gaussian modelling that focuses on accurate estimation of the tails of the HM distribution.}

\begin{figure}[h]
    \centering
     \includegraphics[scale=0.3]{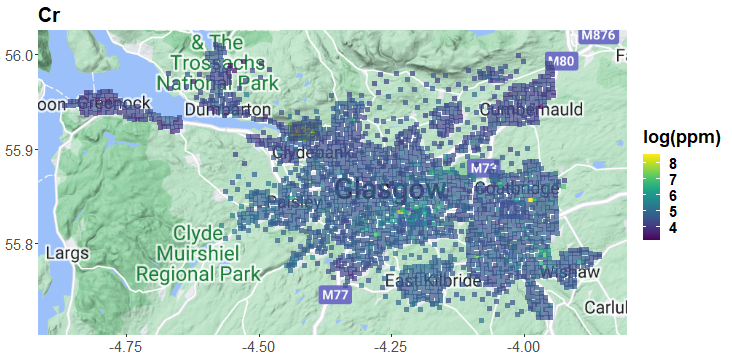}
     \includegraphics[scale=0.3]{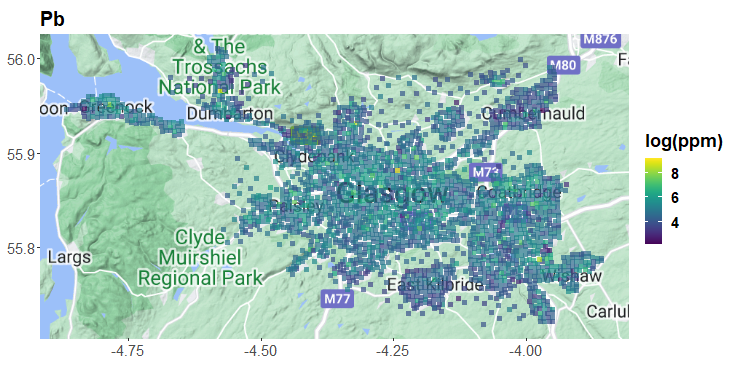}
    \includegraphics[scale=0.3]{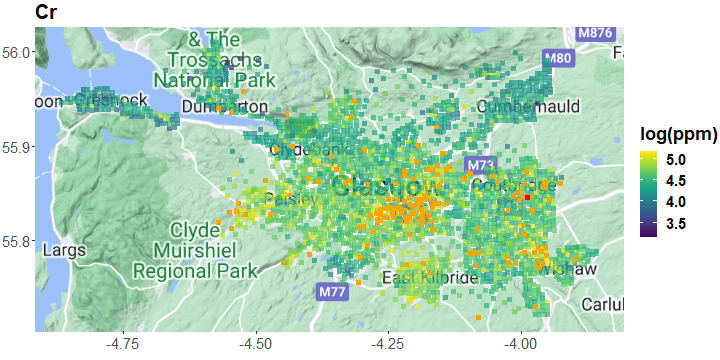}
    \includegraphics[scale=0.3]{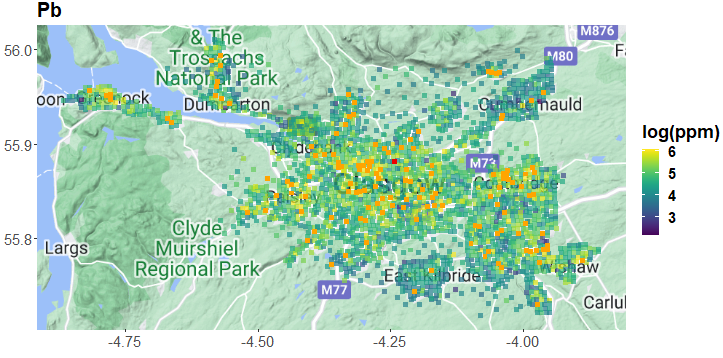}\\
    \vspace{-1mm}
    \begin{flushleft}
     \hspace*{9mm} {\tiny{{Soils Data BGS \copyright{}  NERC. Map data \copyright{} 2023 Google.}}}
    \end{flushleft}
    
    \caption{Top: Maps of the log concentrations of Chromium (left) and Lead (right) in the Glasgow Conurbation.Bottom: Censored map of log concentrations where observations above the $95$th percentile are orange ($5.198$ for Cr and $6.095$ for Pb), and the maximum observation is marked in red ($8.582$ for Cr and $9.204$ for Pb).}
    
    \label{fig:CrPbMap}
\end{figure}

{Maps of the concentrations of Cr and Pb can be seen in Figure \ref{fig:CrPbMap}. The top row shows the full range of concentrations of Cr and Pb after a log transformation, respectively. The bottom row} maps display a continuous scale of observations up to the 95th quantile. The observations exceeding this threshold are in solid orange colour and the maximum value is shown in red. In this scale, the 95th quantile {corresponds to $5.198$ log(ppm) for Cr and $6.095$ log(ppm) for Pb, while the maximum values are $8.582$ log(ppm) and $9.204$ log(ppm) for Cr and Pb respectively.} The map shows an agglomeration of high values just south of the Clyde in central Glasgow, while other high values can also be found in Coatbridge, East Kilbride, and Wishaw to the south and southeast. To the southwest, high values are seen around Paisley and further south towards Clyde Muirshiel Reginal Park. High values for Pb are found near Greenock Port and Dumbarton, especially along the  M74 and the A82 towards The Trossachs National Park and the port of Greenock.
% \dcc{for completeness, include 95th quantile and max of Pb}
% \boldarial{hello}

Following \citet{johnson_modelling_2017}, we obtained terrain and topography variables for model covariates such as elevation, slope, aspect, multiresolution index of valley bottom flatness (MRVBF), the complementary multiresolution index of the ridge top flatness (MRRTF), and topographic wetness index (TWI) from a digital elevation model (DEM) by the Ordinance Survey (OS) at a resolution of 1:50000. Processing of the data was done in \texttt{R} using \texttt{RSAGA} \citep{brenning_statistical_2008}. Variables providing proximal traffic information such as the type of nearest road, primary (A) or secondary (B), and distance to the nearest primary and secondary roads were obtained from the UK Department of Environment, Food, and Agricultural Affairs (DEFRA). The covariates were introduced in the linear predictor as linear fixed effects. 

\subsection{Results}
\begin{figure}[h]
    \centering
    \includegraphics[scale=0.3]{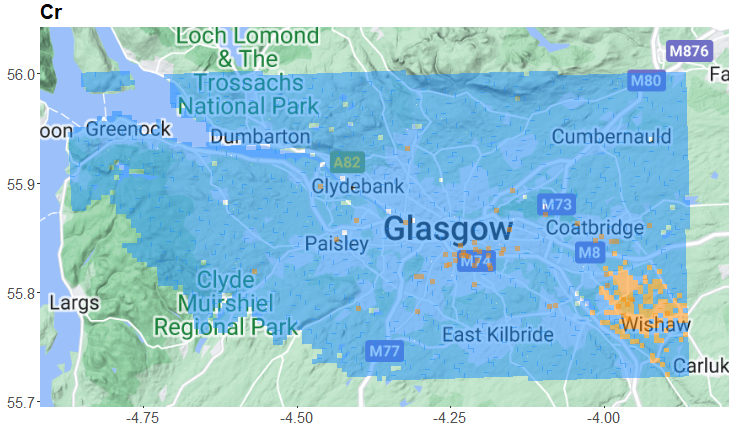}
    \includegraphics[scale=0.3]{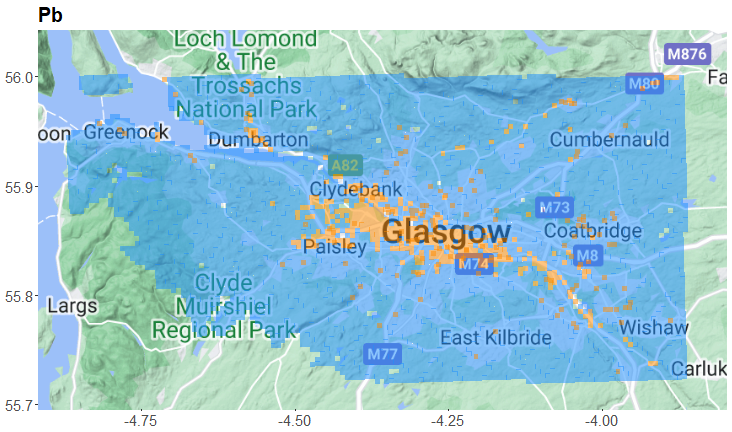}
     \vspace{-1mm}
    \begin{flushleft}
     \hspace*{8mm} {\tiny{{Soils Data BGS \copyright{}  NERC. Map data \copyright{} 2023 Google.}}}
    \end{flushleft}
    \caption{Classification map of observations for Cr (left) and Pb (right) {as body (blue) or tail (orange)} using the mixture proportions $p_1=0.98$ and $p_2=0.95$ respectively.}
    \label{fig:binom_map}
\end{figure}

The \textit{a priori} classification of observations as body or tail depends on the parameters $p_1$ and $p_2$, corresponding to the mixture proportions of $y_1$ and $y_2$ respectively, which indicate the probability of an observation belonging to the body of the distribution. Unlike the simulation study, {$p_1$ and $p_2$ are} not known; therefore,  an exhaustive search for appropriate values is performed using a grid of values from $0.75$ to $0.99$ in increments of $0.01$, and the values for $p_1$ and $p_2$ that yield the best DIC and predictive RMSE values are selected. In the case of the Cr-Pb pair, the best model performance was given by $p_1=0.98$ and $p_2=0.95$. A prediction using a spatial binomial model is made over a rectangular region representing the spatial extent of the observations in order to obtain the classification of prediction locations
% ,\dcc{is there something missing here?} 
(Figure \ref{fig:binom_map}). The binomial model captures the pattern of extreme observations in Figure \ref{fig:CrPbMap}, where extreme concentrations of Cr are clustered south of the Clyde in Glasgow and Wishaw. {The predictions} for Pb also capture the true pattern, with extreme or tail observations found along the M74, the A82, and Coatbridge.
% \dcc{why so much empty space around figure?}

\begin{figure}[h]
    \centering
    \includegraphics[scale=0.3]{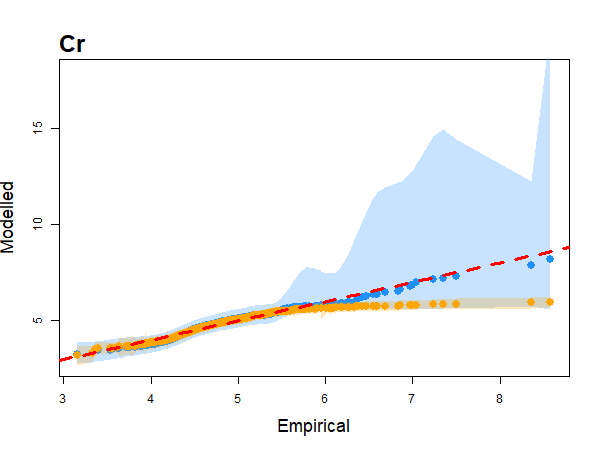}
    % {figures/QQ_Var1+SE_MANUNEW2.png}
    \includegraphics[scale=0.3]{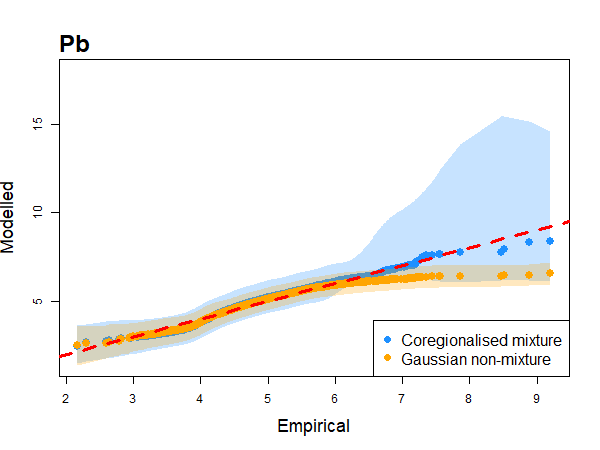}
    % {figures/QQ_Var2+SE_MANUNEW2.png}
    \caption{Q-Q plots for the coregionalised mixture model (blue) and smoothed credible intervals for the coregionalised mixture model predictions of the log concentrations of Cr and Pb when the proportions are $p_1=0.98$ and $p_2=0.95$ and the Gaussian model fitted in INLA (orange).}
    \label{fig:qq_crpb}
\end{figure}

{Our bivariate spatial extreme mixture model in \eqref{eq:bivmodel}} is fitted to the rectangular area representing the spatial extent of the observations to obtain a continuous prediction of the concentrations of Cr and Pb in the river basin. Model validation is performed using k-fold cross-validation, with $k=20$ where every fold removes $5\%$ of the observations and performs predictions on the removed locations. Figure \ref{fig:qq_crpb} compares the cross-validation predictions of the coregionalised mixture model to a non-mixture Gaussian model fitted in INLA where both components are fitted independently of each other {and provides the smoothed $95\%$ credible intervals. The larger width of the credible intervals of the coregionalised mixture model is expected due to the heavier tail of the GPD distribution compared to the Gaussian model. For Cr, we see the deviation is greater, which is understandable given the heavier tail of the Cr distribution as corroborated by its estimated kurtosis. On the other hand, the difference between models is less distinct for Pb and can be explained due to lower kurtosis. Overall, the coregionalised mixture model shows an improvement over the Gaussian model in capturing extreme values and modelling a heavy-tailed distributions.}

\begin{figure}[H]
    \centering
     \includegraphics[scale=0.3]{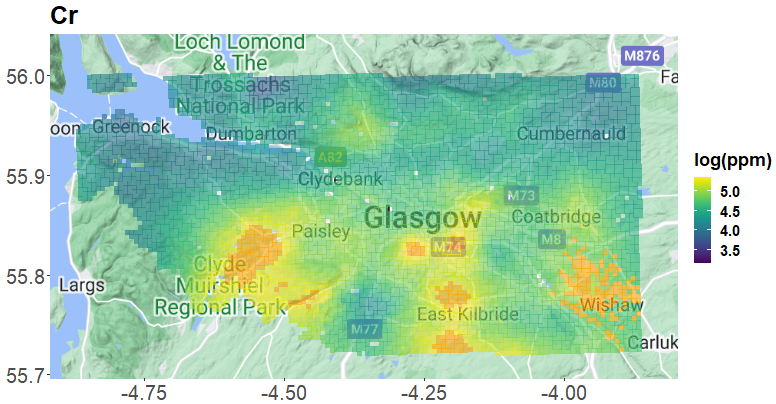}
    \includegraphics[scale=0.3]{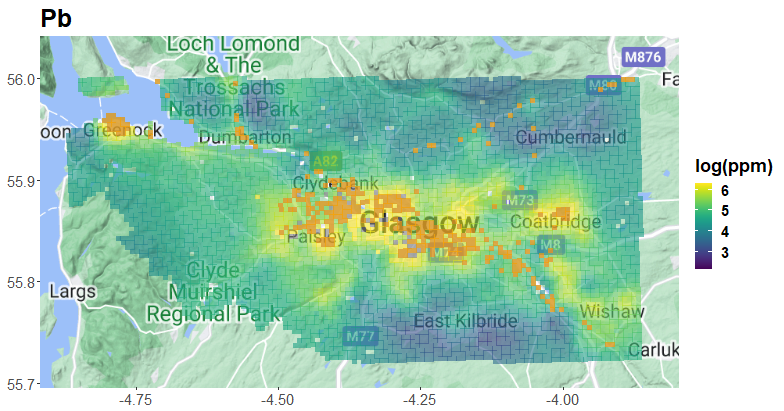}
     \includegraphics[scale=0.3]{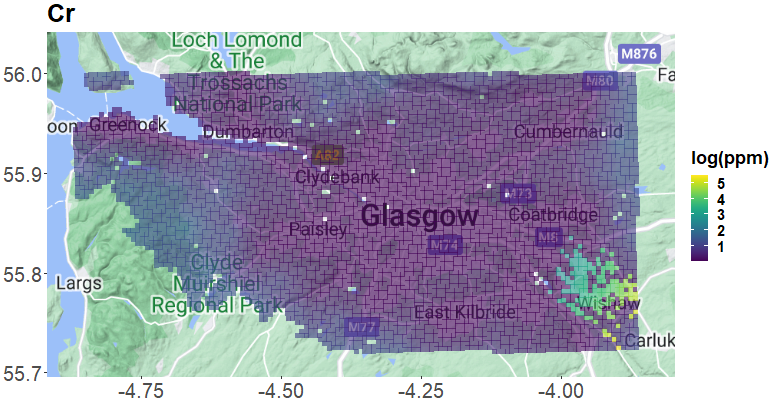}
    \includegraphics[scale=0.3]{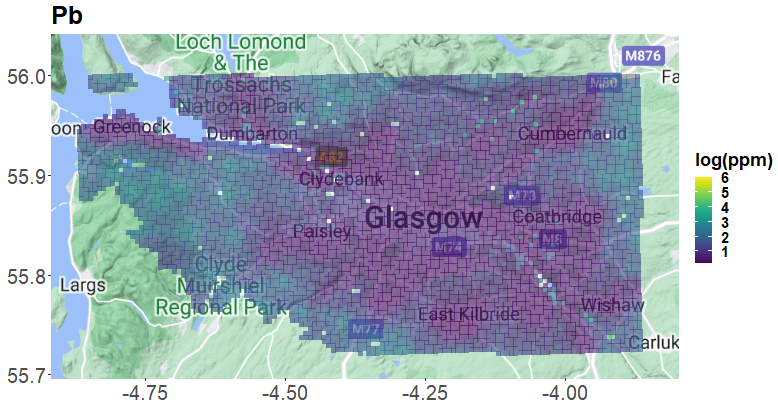}
     \vspace{-1mm}
    \begin{flushleft}
     \hspace*{5mm} {\tiny{{Soils Data BGS \copyright{}  NERC. Map data \copyright{} 2023 Google.}}}
    \end{flushleft}
    \caption{Top: maps of the results of the coregionalised mixture model with observations exceeding the 95th quantile in colour orange (same values as in Figure \ref{fig:CrPbMap}) for Cr (left) and Pb (right). Bottom: width of the {$95\%$ credible} intervals for the coregionalised mixture model for both contaminants.}
    \label{fig:mix_map}
\end{figure}

The results given in Figure \ref{fig:mix_map} {provide maps of the model results as marginal concentrations. The figures} show that Cr has new predicted areas of contamination in the Wishaw area to the southeast. The area between the city centre and East Kilbride to the south experiences higher concentrations too, which match the observed data. Other areas of raised predicted Cr concentrations are west of Paisley to the west of Glasgow, and Coatbrige to the east. Pb shows similar trends to those anticipated. The M74 road around the city centre and to the south through Wishaw are singled out as having especially high concentrations, as does the Port of Greenock and the A82 towards the Trossachs National Park. Additional predicted areas of contamination include the M80 near Falkirk to the southeast.  Overall, higher concentrations can be found along more densely populated areas from Paisley to the west to Coatbridge in the east and around major A and M roads with heavy traffic. 

\begin{figure}[h]
    \centering
    \includegraphics[scale=0.3]{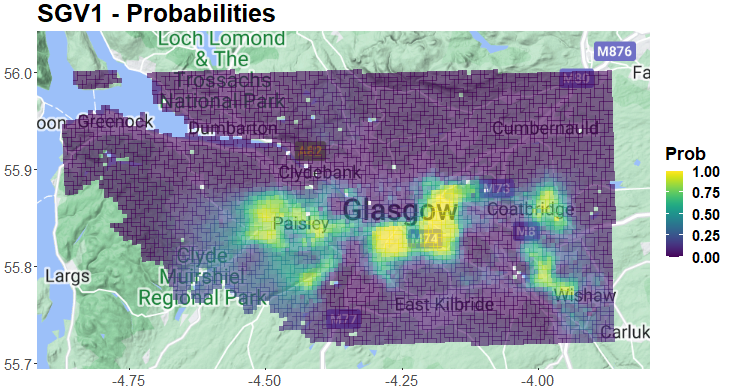}
    \includegraphics[scale=0.3]{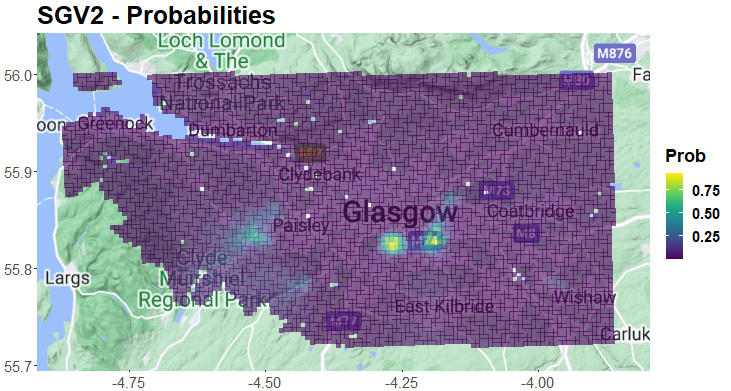}
    \includegraphics[scale=0.3]{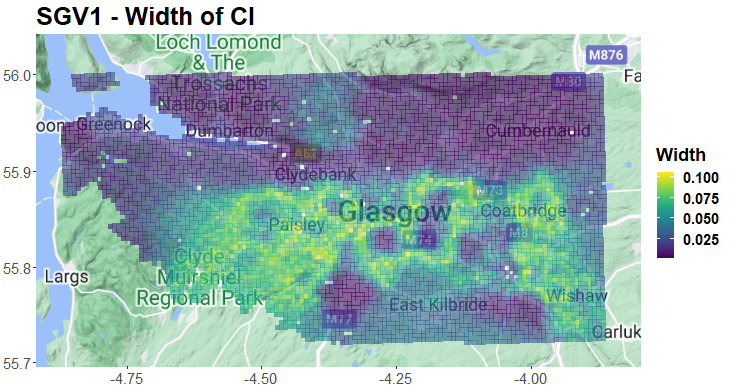}
    \includegraphics[scale=0.3]{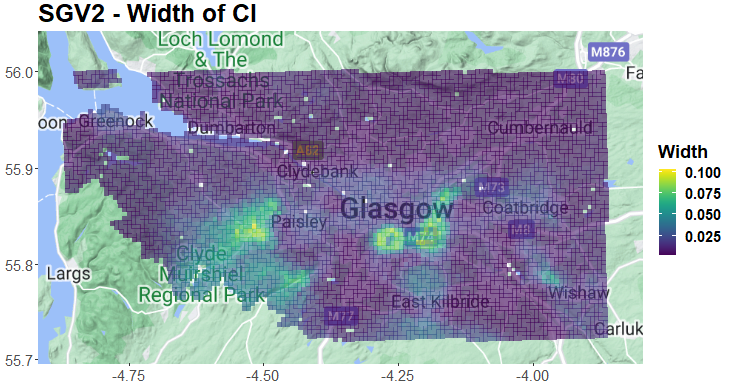}
    \includegraphics[scale=0.3]{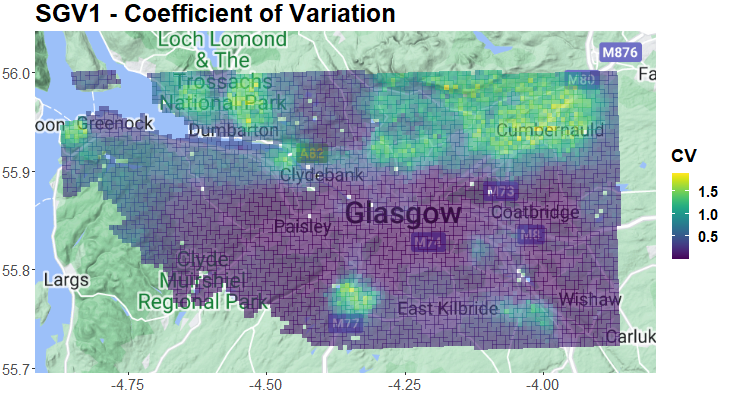}
    \includegraphics[scale=0.3]{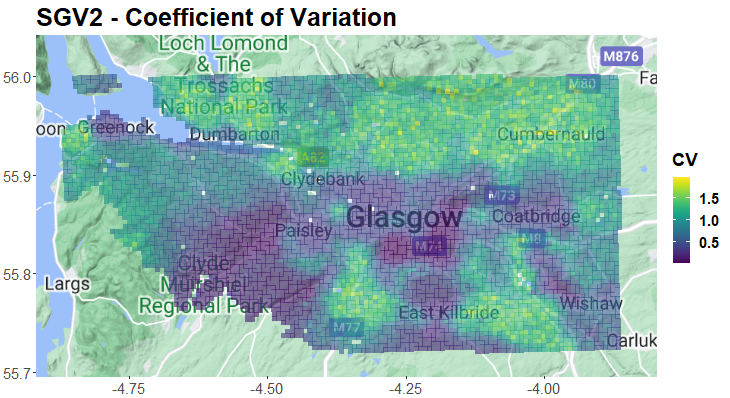}
     \vspace{-1mm}
    \begin{flushleft}
     \hspace*{8mm} {\tiny{{Soils Data BGS \copyright{}  NERC. Map data \copyright{} 2023 Google.}}}
    \end{flushleft}
    \caption{{Top: Probability maps for joint exceedances SGV1 (left) and SGV2 (right). Middle: Width of $95\%$ confidence intervals of the exceedance probabilities as described in Section \ref{sec:MixCoreg} for SGV1 and SGV2, respectively. Bottom: Maps of the coefficient of variation for the estimated exceedance probabilities.}}
    \label{fig:prob_maps}
\end{figure}

{Assessing the risk posed by both contaminants simultaneously is possible using joint exceedance probability maps. The Environment Agency in the UK has published SGV for most HM elements \citep{cole_using_2009}, indicating what concentration thresholds are considered safe. In residential areas with no plants or food production (SGV1), Cr is recommended to stay under 130ppm, 4.87 in log(ppm), and Pb under 200ppm, 5.30 log(ppm). In residential areas with plant or food production (SGV2), the SGV is 200ppm for Cr and 310ppm for Pb, or 5.29 log(ppm) and 5.74 log(ppm) for Cr and Pb, respectively. Using the process described in Figure \ref{fig:mc_probs}, we computed the probabilities of joint exceedance of SGV1 and SGV2 using samples from the posterior predictive distribution. The maps in Figure \ref{fig:prob_maps} show that Cr and Pb have high probabilities of exceeding SGV1 in the south, southeast, and east of the city of Glasgow. These areas are well-known for legacy contamination due to historical chromium ore processing and other chromium-producing industries \citep{CLAIRE_2007}. The figures also show that the uncertainties given by the confidence intervals are small and do not affect interpretation. The coefficient of variation, defined as the ratio of the standard deviation to the mean, shows that areas of high probability exhibit small relative variation, whereas the areas of low probability, mainly to the north of the city, show larger variation in relation to the mean. }

{The maps for SGV2 show a similar result where there is only a high probability of both contaminants exceeding the threshold in two contamination hotspots in well-known areas to the south and southeast of the city centre. Additionally, there are higher probabilities to the southwest, near Paisley. Uncertainty maps show the two hotspots of contamination with high certainty, while contamination detected to the southwest is more uncertain. Areas of low probability, such as the north, show greater variability in relation to the mean, similar to SGV1. }

\section{Conclusion and Discussion}\label{sec:C&D}
Maps of geochemical concentrations are generally produced using geostatistical models under a Gaussian framework. Such models do not account for the various processes responsible for the spatial distribution of geochemical concentrations but rather model all observations as realisations of a single Gaussian process, resulting in an underestimation of extreme values and measures of risk . When more than one contaminant is present, classical geostatistical models not only under-estimate extreme concentrations, but also do not account for the extremal dependence between contaminants. We propose to partition the distribution of each contaminant into body and tail, representing non-extreme and extreme concentrations, and weaving them together inside a coregionalised mixture model framework. The body component, representing non-extreme observations linked to a combination of natural pedogenic processes and diffuse background contamination, is modelled using a Gaussian model common to geochemical applications. The tail, containing extreme observations related to contaminating anthropogenic or natural processes, is modelled using an adapted extreme value distribution. While EVT is an attractive framework to capture extremal behaviour, it requires replications at each location, which are not commonly available in geochemical datasets. For this reason, a transformation is applied to the tail under a stationary GPD, and later modelled under a Gaussian likelihood following the usual geostatistical setting of a single replication. Our proposed framework combines ideas from coregionalisation models, mixture models, spatial latent Gaussian models, and EVT to capture non-extremal concentrations as well as the extremal behaviour and dependence between two contaminants at high concentrations and produce a continuous map of concentrations.  

We fit our model to Cr and Pb concentrations in the Glasgow Conurbation in the west of Scotland. {The data was first transformed using a log transformation, and kurtosis provided evidence of the deviation of the tails from a Gaussian distribution}.
% \dcuba{We suggest assessing the tail of the distribution using kurtosis as a pre-processing step. If the distribution is within the expected kurtosis values of a Gaussian distribution for the sample size, the data could be considered Gaussian and modelled as such. When the kurtosis of the distribution provides evidence of a heavier tail, a more appropriate model, such as the coregionalised mixture model proposed here, can be used}. 
The results of the model show that there are areas of Cr contamination to the south and southwest of Glasgow, southwest of Paisley, to the south around East Kilbride, and to the east near Wishaw. Areas of high Pb concentrations are found around the Glasgow city centre and seem to be contained to areas around the Clyde River and the Port of Greenock. Marginal credible intervals are the widest for observations belonging to the tail, which is expected for extreme observations due to the limited sample size. {The model shows that joint contamination has a high probability of exceeding SGV1 and SGV2 in the south and southeast of the city of Glasgow, areas known to be affected by legacy industrial HM contamination}. A comparison with a classical Gaussian model shows that the joint modelling of two contaminants using the coregionalised mixture model is an improvement, particularly at the tail. Furthermore, the {shared spatial random component $z_{T_1}$ models the factors affecting the extreme values of both contaminants while the weight of the shared spatial component, $\lambda$, models the dependence between both tail distributions. High values of $\lambda$ can account for strong extremal dependence between components while $\lambda = 0$ indicates extremal independence.}. For this application, we constrained the dependence structure to only account for extremal dependence through a {shared} spatial random effect. However, the framework is flexible and can be easily extended to capture dependence through other terms in both the body and tail or extended beyond the bivariate case. Future work can be developed to jointly model $p_1$ and $p_2$ in space and integrate this modelling with the coregionalised mixture model. A natural way to do this is through a hierarchical Bayesian model where inference is carried out using simulation-based approaches, such as MCMC.
\newpage

\section*{Appendix}
\begin{figure}[hbt!]
    \centering
    \includegraphics[scale=0.3]{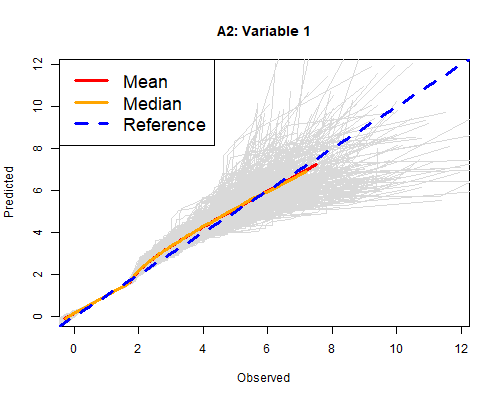}
    \includegraphics[scale=0.3]{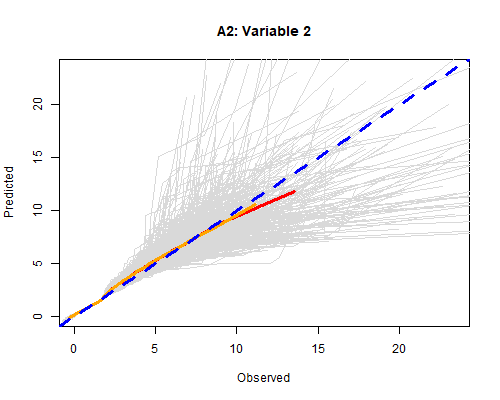}
    \caption{Q-Q plots of all simulations (grey) for simulation study A2, for variables 1 and 2. The mean and median of the simulations are shown in red and orange, respectively, while the reference line is in blue. }
    \label{fig:resultsA2}
\end{figure}

\begin{table}[hbt!]
\centering
\begin{tabular}{rlrrrrrr}
  \hline
  &\textbf{Parameter} &\textbf{True Val}& \textbf{Median} & \textbf{Mean} & \textbf{Sd} & \textbf{Coverage.pr} & \textbf{RMSE} \\
  \hline
 & $\alpha_{B1}$ & 1.00 & 1.02 & 1.02 & 0.01 & 0.99 & 0.03 \\ 
   & $\alpha_{T1}$ & 0.00 & -0.12 & -0.11 & 0.10 & 0.96 & 0.15 \\ 
   & $\alpha_{B2}$ & 1.00 & 1.03 & 1.02 & 0.01 & 0.99 & 0.03 \\ 
   & $\alpha_{T2}$ & 0.00 & -0.12 & -0.11 & 0.10 & 0.99 & 0.16 \\ 
   & $\beta_{B1_{1}}$ & 0.10 & 0.09 & 0.09 & 0.01 & 0.99 & 0.01 \\ 
   & $\beta_{B1_{2}}$ & 0.25 & 0.22 & 0.22 & 0.01 & 0.99 & 0.03 \\ 
   & $\beta_{T1_{1}}$ & 0.10 & 0.04 & 0.04 & 0.10 & 0.97 & 0.12 \\ 
   & $\beta_{T1_{2}}$ & 0.25 & 0.07 & 0.06 & 0.08 & 0.64 & 0.20 \\ 
   & $\beta_{B2_{1}}$ & 0.10 & 0.09 & 0.09 & 0.01 & 0.99 & 0.01 \\ 
   & $\beta_{B2_{2}}$ & 0.25 & 0.22 & 0.22 & 0.01 & 0.99 & 0.03 \\ 
   & $\beta_{T2_{1}}$ & 0.10 & 0.04 & 0.04 & 0.11 & 0.94 & 0.12 \\ 
   & $\beta_{T2_{2}}$ & 0.25 & 0.08 & 0.07 & 0.09 & 0.70 & 0.19 \\ 
   & $\tau_{1}$ & 0.01 & 0.33 & 0.25 & 0.23 & 0.60 & 0.40 \\ 
   & $\tau_{2}$ & 0.01 & 0.21 & 0.15 & 0.17 & 0.90 & 0.26 \\ 
   & $\rho_{T1}$ & 10.00 & 55.68 & 34.38 & 91.22 & 0.84 & 101.43 \\ 
   & $\rho_{T2}$ & 15.00 & 112.76 & 99.45 & 69.51 & 0.74 & 119.67 \\ 
   & $\lambda$\ & 0.25 & 0.97 & 0.98 & 0.12 & 0.20 & 0.73 \\ 
   & $\sigma_1$ & 1 & 1.04 & 1.02 & 0.24 & 0.94 & 0.02\\
   & $\sigma_2$ & 1 & 1.22 & 1.19 & 0.30 & 0.93 & 0.01\\
   & $\xi_1$\ & 0.05 & 0.03 & 0.03 & 0.17 & 0.92 & 0.003 \\ 
   & $\xi_2$\ & 0.25 & 0.21 & 0.20 & 0.19 & 0.91 & 0.002 \\ 
   \hline\\
\end{tabular}
\caption{Summary of results for simulation study A2. The table shows the parameter's true value; estimated parameter mean, median and standard deviation; the 95\% coverage probability; and the mean RMSE.}
\label{tab:A2}
\end{table}

\begin{table}[hbt!]
\centering
\begin{tabular}{cccccc}
\hline
\textbf{Scenario}              & \textbf{Variable} & \textbf{Accuracy} & \textbf{Precision} & \textbf{Sensitivity} & \textbf{Specificity} \\ \hline
\multirow{2}{*}{A1} &  1        & 0.89              & 0.95               & 0.58                 & 0.99                 \\
                               &  2        & 0.89              & 0.96               & 0.57                 & 0.99                \\
\multirow{2}{*}{A2} &  1        & 0.91              & 0.77               & 0.62                 & 0.97                 \\
                               &  2        & 0.91              & 0.78               & 0.62                 & 0.98                 \\
                               \hline\\
% \multirow{2}{*}{Scenario C: 3} & Variable 1        & 0.80              & 0.60               & 0.63                 & 0.85                 \\
                               % & Variable 2        & 0.71              & 0.44               & 0.49                 & 0.78                 \\ \hline
\end{tabular}

\caption{Classification evaluation of simulation scenarios A1 and A2.}
\label{tab:allA}
\end{table}

\begin{figure}[hbt!]
    \centering
    \includegraphics[scale=0.3]{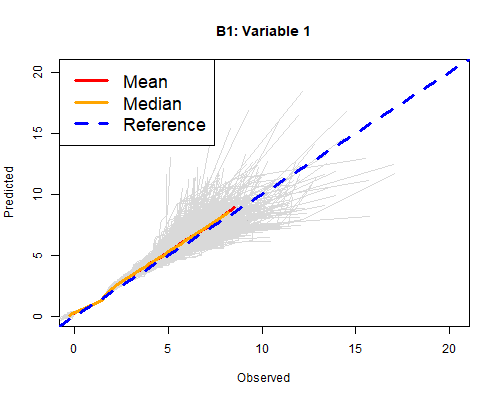}
    \includegraphics[scale=0.3]{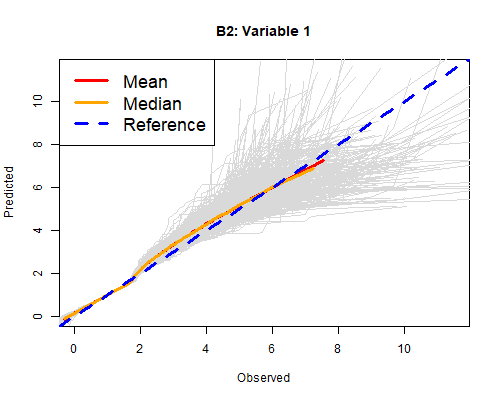}
    \caption{Q-Q plots of simulations (grey) for simulation scenario B1, for variables 1 and 2. The mean and median of the simulations are shown in red and orange, respectively, while the reference line is given in blue. }
    \label{fig:resultsB1}
\end{figure}

\begin{table}[hbt!]
\centering
\begin{tabular}{rlrrrrrr}
  \hline
 &\textbf{Parameter} &\textbf{True Val}& \textbf{Median} & \textbf{Mean} & \textbf{Sd} & \textbf{Coverage.pr} & \textbf{RMSE} \\
  \hline
 & $\alpha_{B1}$ & 1.00 & 1.04 & 1.04 & 0.01 & 0.99 & 0.04 \\ 
   & $\alpha_{T1}$ & 0.00 & -0.22 & -0.22 & 0.06 & 0.42 & 0.23 \\ 
   & $\alpha_{B2}$ & 1.00 & 1.05 & 1.04 & 0.03 & 0.99 & 0.06 \\ 
   & $\alpha_{T2}$ & 0.00 & -0.22 & -0.21 & 0.06 & 0.93 & 0.22 \\ 
   & $\beta_{B1_{1}}$ & 0.10 & 0.07 & 0.07 & 0.01 & 0.99 & 0.03 \\ 
   & $\beta_{B1_{2}}$ & 0.25 & 0.18 & 0.18 & 0.01 & 0.99 & 0.07 \\ 
   & $\beta_{T1_{1}}$ & 0.10 & 0.06 & 0.06 & 0.06 & 0.97 & 0.08 \\ 
   & $\beta_{T1_{2}}$ & 0.25 & 0.17 & 0.17 & 0.05 & 0.97 & 0.09 \\ 
   & $\beta_{B2_{1}}$ & 0.10 & 0.08 & 0.07 & 0.01 & 0.99 & 0.03 \\ 
   & $\beta_{B2_{2}}$ & 0.25 & 0.19 & 0.19 & 0.02 & 0.99 & 0.06 \\ 
   & $\beta_{T2_{1}}$ & 0.10 & 0.07 & 0.07 & 0.06 & 0.97 & 0.07 \\ 
   & $\beta_{T2_{2}}$ & 0.25 & 0.16 & 0.17 & 0.05 & 0.96 & 0.10 \\ 
   & $\tau_{1}$ & 0.01 & 0.22 & 0.21 & 0.10 & 0.64 & 0.23 \\ 
   & $\tau_{2}$ & 0.01 & 0.16 & 0.14 & 0.07 & 0.91 & 0.16 \\ 
   & $\rho_{T1}$ & 10.00 & 55.54 & 26.09 & 301.62 & 0.93 & 304.05 \\ 
   & $\rho_{T2}$ & 15.00 & 107.33 & 82.27 & 69.61 & 0.81 & 115.48 \\ 
   & $\lambda$\ & 0.90 & 0.98 & 0.98 & 0.10 & 0.99 & 0.13 \\ 
   & $\sigma_1$ & 1 & 0.92 & 0.91 & 0.13 & 0.82 & 0.15\\
   & $\sigma_2$ & 1 & 1.02 & 1.01 & 0.15 & 0.93 & 0.15\\
   & $\xi_1$\ & 0.05 & 0.09 & 0.1 & 0.11 & 0.92 & 0.02 \\ 
   & $\xi_2$\ & 0.25 & 0.29 & 0.29 & 0.11 & 0.96 & 0.05 \\ 
   \hline
   \\
\end{tabular}
\caption{Summary of results for simulation scenario B1. The table shows the parameter's true value; estimated parameter mean, median and standard deviation; the 95\% coverage probability; and the mean RMSE.}
\label{tab:B1}
\end{table}

\begin{figure}[hbt!]
    \centering
    \includegraphics[scale=0.3]{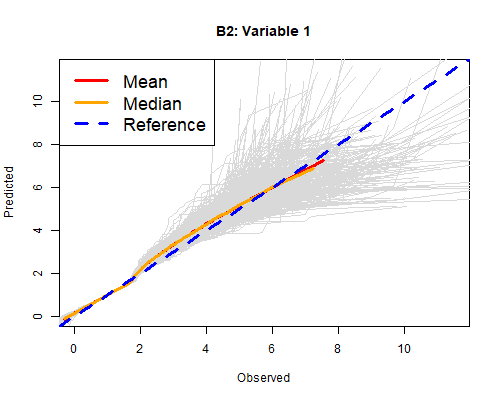}
    \includegraphics[scale=0.3]{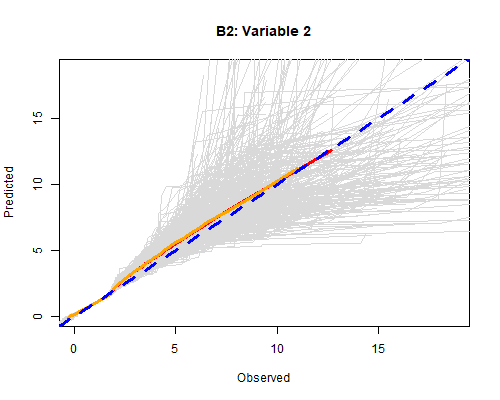}
    \caption{Q-Q plots of all simulations of scenario B2, for variables 1 and 2. The mean and median of the simulations are shown in red and orange, respectively, while the reference line is given in blue. }
    \label{fig:resultsB2}
\end{figure}

\begin{table}[hbt!]
\centering
\begin{tabular}{rlrrrrrr}
  \hline
 &\textbf{Parameter} &\textbf{True Val}& \textbf{Median} & \textbf{Mean} & \textbf{Sd} & \textbf{Coverage.pr} & \textbf{RMSE} \\
  \hline
 & $\alpha_{B1}$ & 1.00 & 1.02 & 1.02 & 0.01 & 0.99 & 0.03 \\ 
   & $\alpha_{T1}$ & 0.00 & -0.11 & -0.11 & 0.10 & 0.94 & 0.15 \\ 
   & $\alpha_{B2}$ & 1.00 & 1.02 & 1.02 & 0.01 & 0.99 & 0.03 \\ 
   & $\alpha_{T2}$ & 0.00 & -0.09 & -0.09 & 0.10 & 0.99 & 0.13 \\ 
   & $\beta_{B1_{1}}$ & 0.10 & 0.09 & 0.09 & 0.01 & 0.99 & 0.01 \\ 
   & $\beta_{B1_{2}}$ & 0.25 & 0.22 & 0.22 & 0.01 & 0.99 & 0.03 \\ 
   & $\beta_{T1_{1}}$ & 0.10 & 0.03 & 0.03 & 0.11 & 0.94 & 0.13 \\ 
   & $\beta_{T1_{2}}$ & 0.25 & 0.07 & 0.07 & 0.08 & 0.64 & 0.20 \\ 
   & $\beta_{B2_{1}}$ & 0.10 & 0.09 & 0.09 & 0.01 & 0.99 & 0.01 \\ 
   & $\beta_{B2_{2}}$ & 0.25 & 0.22 & 0.22 & 0.01 & 0.99 & 0.03 \\ 
   & $\beta_{T2_{1}}$ & 0.10 & 0.03 & 0.03 & 0.11 & 0.92 & 0.13 \\ 
   & $\beta_{T2_{2}}$ & 0.25 & 0.06 & 0.06 & 0.08 & 0.62 & 0.20 \\ 
   & $\tau_{1}$ & 0.01 & 0.34 & 0.22 & 0.27 & 0.66 & 0.42 \\ 
   & $\tau_{2}$ & 0.01 & 0.23 & 0.17 & 0.20 & 0.89 & 0.29 \\ 
   & $\rho_{T1}$ & 10.00 & 45.81 & 32.84 & 65.71 & 0.81 & 74.65 \\ 
   & $\rho_{T2}$ & 15.00 & 151.84 & 105.32 & 271.00 & 0.76 & 302.83 \\ 
   & $\lambda$\ & 0.90 & 0.95 & 0.99 & 0.17 & 0.97 & 0.72 \\ 
      & $\sigma_1$ & 1 & 1.06 & 1.03 & 0.25 & 0.91 & 0.19\\
   & $\sigma_2$ & 1 & 1.20 & 1.17 & 0.30 & 0.92 & 0.2\\
   & $\xi_1$\ & 0.05 & 0.018 & 0.03 & 0.19 & 0.88 & 0.18 \\ 
   & $\xi_2$\ & 0.25 & 0.223 & 0.23 & 0.20 & 0.90 & 0.20 \\ 
   \hline
   \\
\end{tabular}

\caption{Summary of results for simulation scenario B2. The table shows the parameter's true value; estimated parameter mean, median and standard deviation; the 95\% coverage probability; and the mean RMSE.}
\label{tab:B2}
\end{table}

\begin{table}[hbt!]
\centering
\begin{tabular}{cccccc}
\hline
\textbf{Scenario}              & \textbf{Variable} & \textbf{Accuracy} & \textbf{Precision} & \textbf{Sensitivity} & \textbf{Specificity} \\ \hline
\multirow{2}{*}{B1} &  1        & 0.88              & 0.92               & 0.55                 & 0.98                 \\
                               &  2        & 0.88              & 0.93               & 0.56                 & 0.99                \\
\multirow{2}{*}{B2} &  1        & 0.93              & 0.61               & 0.91                 & 0.93                 \\
                               &  2        & 0.93              & 0.61               & 0.91                 & 0.93                 \\
                               \hline\\
% \multirow{2}{*}{Scenario C: 3} & Variable 1        & 0.80              & 0.60               & 0.63                 & 0.85                 \\
                               % & Variable 2        & 0.71              & 0.44               & 0.49                 & 0.78                 \\ \hline
\end{tabular}

\caption{Classification evaluation of scenarios B1 and B2.}
\label{tab:allB}
\end{table}

\clearpage
\section*{Declarations}
\bmhead{Ethical Approval} 
Not Applicable.

\bmhead{Availability of supporting data}
The GBASE dataset can be obtained from the British Geological Survey (enquiries@bgs.ac.uk).

\bmhead{Competing interests}
Not Applicable.

\bmhead{Funding} 

\bmhead{Authors' contributions}
M.D.C. developed the methodology presented under the supervision of D.C.C., M.S., and B.M. M.D.C. wrote the draft, receiving editing feedback from D.C.C., M.S., and B.M. 

% \bmhead{Acknowledgments}

\newpage
\bibliography{sn}
\end{document}